\documentclass[aip,jcp,groupedaddress,reprint,nofooinbib]{revtex4-1}

\usepackage{graphicx}
\usepackage{graphics}
\usepackage{amsfonts}
\usepackage{amsmath}
\usepackage{amssymb}
\usepackage{color}
\usepackage{soul}
\usepackage{multirow}
\usepackage{chngcntr}

\newcommand{\bracketed}[1]{\left\langle{#1}\right\rangle}

\setcitestyle{super}

\begin{document}

\title{Lattice simulation method to model diffusion and NMR spectra in porous materials}

\author{C\'eline Merlet$^{1,*}$}
\author{Alexander C. Forse$^1$}
\author{John M. Griffin$^{1}$}
\author{Daan Frenkel$^{1}$}
\author{Clare P. Grey$^{1}$}
\affiliation{$^1$ \small Department of Chemistry, University of Cambridge, Lensfield Road, Cambridge CB2 1EW, UK}
\affiliation{$^*$ \small Corresponding author: celine.merlet.fr@gmail.com}

\date{\today}

\begin{abstract}

A coarse-grained simulation method to predict NMR spectra of ions diffusing in porous carbons is proposed. The coarse-grained model uses input from molecular dynamics simulations such as the free-energy profile for ionic adsorption, and density-functional theory calculations are used to predict the NMR chemical shift of the diffusing ions. The approach is used to compute NMR spectra of ions in slit pores with pore widths ranging from 2 to 10~nm. As diffusion inside pores is fast, the NMR spectrum of an ion trapped in a single mesopore will be a sharp peak with a pore size dependent chemical shift. To account for the experimentally observed NMR line shapes, our simulations must model the relatively slow exchange between different pores. We show that the computed NMR line shapes depend on both the pore size distribution and the spatial arrangement of the pores. The technique presented in this work provides a tool to extract information about the spatial distribution of pore sizes from NMR spectra. Such information is difficult to obtain from other characterisation techniques.\\

\end{abstract}


\maketitle

\newpage

\section{Introduction}
\label{Intro}

Transport in porous materials is relevant for phenomena as diverse as energy conversion and storage, gas storage, heterogeneous catalysis, cell growth and drug delivery~\cite{Zhao06}. In all these cases, the performance of the porous material depends on its structural properties, the latter determining the geometry and dynamical properties of the confined fluid. The characterisation of porosity and the understanding of its effect on performance are therefore essential to optimise the porous materials for their respective applications. Nuclear Magnetic Resonance (NMR) is an experimental method that is uniquely suited to probe both microscopic structure and the dynamical properties of fluids confined in porous media.

NMR has been widely used to investigate the adsorption of various fluids in porous carbons~\cite{Xing14,Wang11b,Forse13,Wang13}. Most of these studies report the observation of distinct resonances in the NMR spectra corresponding to adsorbed and freely-diffusing probe species. Irrespective of the nature of the fluid or nucleus studied, the resonances of the species adsorbed on carbon appear at lower frequencies than those of the free species, the difference in the adsorbed and free resonant frequencies depending only weakly on the nature of the fluid. The dominant shift mechanism arises from the magnetic fields caused by the microscopic ring currents of $\pi$ electrons in the graphitic carbon~\cite{Lazzeretti00}. Consistent with this, variations of the structure of the porous carbon have a pronounced effect on the observed NMR resonances~\cite{Xing14,Forse13,Forse14,Borchardt13,Anderson10}.

Although NMR is a powerful technique to investigate both the structure of the carbon and the microscopic structure of the fluid at the carbon surface, the interpretation of the NMR spectra is by no means straightforward as the measured spectra depend in a non-trivial way on the pore structure and on the structure and dynamics of the confined fluid. As a consequence, simplifying assumptions are commonly made to interpret the experimental NMR spectra. 

A necessary (but not sufficient) prerequisite for quantitative interpretation of the NMR spectra is an accurate description of the effect of aromatic ring currents on the local magnetic fields in the porous medium. It is possible to calculate Nucleus Independent Chemical Shifts (NICS) for aromatic molecules using Density Functional Theory (DFT) calculations~\cite{Haddon95,Heine05,Facelli06,Forse14}. These studies provide important insights, accounting, for example, for the shift to lower frequencies of NMR resonances of ions confined in a porous graphitic matrix. However, in order to arrive at a model that can describe all features of the experimental NMR spectra, the effects of local fluid structure and the ionic dynamics must be taken into account. A fully atomistic approach to compute such spectra, let alone an ab-initio calculation, would be prohibitively expensive in view of the wide range of relevant time scales. 

Here, we propose a coarse-grained method that allows us to bridge the gap between molecular simulations and experiment. It is particularly attractive to use lattice models as these allow us to account for the relevant microscopic effects while being computationally much more efficient than off-lattice models. The key computational advantage of the lattice approach used here is that it allows us to sample, in a single simulation, {\em all possible} diffusive trajectories~\cite{Levesque13}, rather than a single one, as would be obtained from an off-lattice simulation or from a conventional lattice simulation of the diffusion of tracers in porous media~\cite{Sen03,Sen04,Dudko05}. The (exponential) computational advantage of this approach turns out to be crucial for exploring the effect of multiple factors on the NMR spectra.  

 
In the following section, we describe the implementation of the lattice gas model and show how it can be used to calculate diffusion coefficients and NMR spectra. We then focus on a two-site exchange model to validate our approach. In particular, we show that we can reproduce the coalescence of NMR resonances observed experimentally. Next, we apply the model to single mesoporous slit pores with different pore widths. Slit pore models have been used extensively in the literature to interpret adsorption isotherms in some classes of porous materials, such as disordered activated carbons for energy storage~\cite{Ravikovitch01,Dash06}, and to simulate ion adsorption and charge storage mechanisms in supercapacitors~\cite{Feng10a,Feng10c,Xing13,Jiang12b}. Their simplicity and wide utilisation make them a good starting point for the application of our model. We describe how the model is parametrised from molecular simulations and explain how diffusion and chemical exchange can lead to the observation of a single chemical shift corresponding to multiple environments. In the last part of this article, we calculate NMR spectra corresponding to ions diffusing in a carbon particle with a realistic pore size distribution and show that the resulting spectra are affected by the spatial distribution of pore sizes. While some experimental techniques, such as adsorption isotherms analysis, can provide the overall pore size distribution, they are not able to probe the spatial distribution of these various pore sizes. As such, the development of appropriate models to predict NMR spectra of species adsorbed in a realistic porous material will open the door to an appropriate interpretation of the NMR experimental data and give insights into both the carbon and liquid structure of the explored systems. This will have implications in various scientific fields where porous materials are used, such as supercapacitors, which will benefit from a detailed characterisation of the electrode/electrolyte interface and the interconnectivity of the pore structure.  

\section{Diffusion coefficients and NMR spectra from a lattice gas model}
\label{Method}

\subsection{Description of the model}

\begin{figure*}[ht!]
\begin{center}
\includegraphics[scale=0.5]{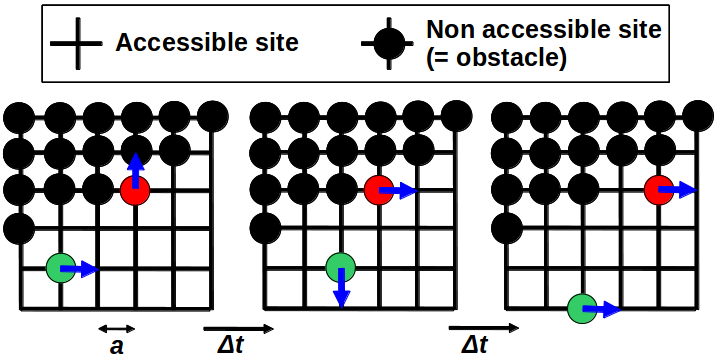}
\end{center}
\caption{Illustration of the lattice gas model. The lattice, characterised by a lattice parameter $a$, is divided between two types of sites: accessible (fluid) and not accessible (porous matrix). If at a given timestep, a particle's velocity points towards an obstacle, then at the next timestep, that particle remains at its original position. The timestep $\Delta t$ relates to the lattice parameter $a$ via the bulk diffusion coefficient $D$~=~$a^2/6\Delta t$ following Eqn.~\ref{eq:D1step}. Note that for the sake of simplicity, the system is represented in two dimensions here but the model is actually three-dimensional.}
\label{model}
\end{figure*}

The simplest lattice-gas models to simulate diffusion describe the diffusing species as non-interacting particles performing kinetic Monte Carlo moves on a lattice that contains both accessible sites (the fluid) and excluded sites (the porous matrix). A schematic representation of such a model is shown in figure~\ref{model}. The lattice model is characterised by a lattice parameter $a$, which is the distance between two lattice nodes, and a timestep $\Delta t$, which corresponds to the typical time it takes the probe particle to diffuse over a distance $a$. The geometry of the matrix is accounted for by the spatial distribution of excluded lattice sites. 

At every timestep, the particles attempt with equal probability to perform a step to any of their nearest neighbours (6 for a three-dimensional simple cubic lattice): if the neighbour site is accessible, the probe particle is moved to the new position, otherwise the particle remains at its original position. We consider three-dimensional cubic lattices with periodic boundary conditions. When modelling diffusion in a porous carbon matrix, we must account for the fact that different accessible sites have different energies (strictly speaking: free energies) and are, therefore, not equally populated. A site energy ($E_i$ for site $i$) is required to account for ionic adsorption at the carbon surface. In addition, when computing NMR lineshapes, we have to account for the fact that the local magnetic field and thus resonant frequency at site $i$, $\nu_i$, will depend on its position. 

To describe diffusion in a spatially varying potential, we vary our Monte Carlo acceptance rules, such that the correct Boltzmann distribution over lattice sites is obtained in equilibrium. Specifically, the conditional probability to accept a Monte Carlo move from site $i$ to site $j$, $p_{acc}(i\rightarrow j)$, which is not necessarily equal to the probability of accepting the reverse move, $p_{acc}(j\rightarrow i)$, is:
\begin{eqnarray}
p_{acc}(i\rightarrow j) & = & \exp \left ( \frac{-(E_j-E_i)}{k_B T} \right ) \hskip 10pt {\rm if} \hskip 5pt E_j > E_i \nonumber \\
 & = & 1 \hskip 10pt {\rm if} \hskip 5pt E_j \leq E_i 
\end{eqnarray}
where $k_B$ is the Boltzmann constant and $T$ is the temperature of the system. As a consequence, the larger $E_j-E_i$ the less likely a particle is to jump from $i$ to $j$. With these rules, the probability of a particle to visit site $i$ will be given by the Boltzmann distribution:
\begin{equation}
\rho_i =<\rho> \times \frac{\exp \left ( \frac{-E_i}{k_B T}\right ) }{\sum_j \exp \left ( \frac{-E_j}{k_B T}\right ) }
\end{equation}  
where $\rho_i$ is the average density at site $i$ and $<\rho>$ is the density averaged over all lattice sites.

\subsection{Diffusion coefficients}   
In MD simulations studies, diffusion coefficients are typically calculated using a Green-Kubo relation that relates the velocity autocorrelation function to the self-diffusion coefficient:
\begin{equation}\label{eq:GK=continuous}
D = \int_0^{\infty} \frac{1}{d}\langle \mathbf{v}(0).\mathbf{v}(t) \rangle .dt 
\end{equation}
where $D$ is the diffusion coefficient, $d$ is the dimensionality of the system, $\mathbf{v}(0)$ is the velocity at time $t=0$, $\mathbf{v}(t)$ is the velocity at time $t$ and $\langle ... \rangle$ denotes an average over all particles and all trajectories. Here we use the discrete equivalent of Eqn.~\ref{eq:GK=continuous} (see Appendix~\ref{app:DIFF}):
\begin{equation}
D = {1\over 2\Delta t} \bracketed{(\Delta x)^2} +{1\over \Delta t}\sum_{j=2}^\infty \bracketed{\Delta x_1\Delta x_j}
\end{equation}
where $\langle ... \rangle$ denotes an average over all considered sites and $\Delta x_i$ denotes the displacement of a particle in the $x$ direction at timestep $i$. If we were to exploit the analogy with off-lattice systems, we would estimate $\bracketed{\Delta x_1\Delta x_j}$ by generating random trajectories for a number of particles starting from different sites in the lattice. This approach is highly inefficient as a very large number of trajectories would have to be generated to obtain good statistics for a heterogeneous system. Here we use a different approach, namely the so-called `moment propagation' method, which was proposed by Frenkel and successfully applied to a number of studies on dynamics of particles in confined systems~\cite{Frenkel87,Rotenberg08,Levesque13}.
 
The moment-propagation method is a recursive scheme that allows us to sample {\em all} possible trajectories of the diffusing particles, rather than a subset. The computational effort scales as $t\times M$, where $t$ is the simulation time and $M$ the number of lattice sites. The computational effort to sample all trajectories in a non-recursive scheme would scale as $z^{t}M$ where $z$ is the coordination number of the lattice. In the expression for diffusion coefficients, the first term is easily calculated as it is equal to $1/2\Delta t$ times the mean-square displacement of a particle in the $x$ direction during one timestep:
\begin{equation} 
{1\over 2\Delta t} \bracketed{(\Delta x)^2}=\frac{1}{2z\Delta t}  \sum_{j=1}^z p_{acc}({i\rightarrow j})a_x^2(j)
\end{equation} 
where $a_x$ is the $x$-component of the vector joining a lattice site to its $j$-th neighbour and the sum runs over all sites $j$ adjacent $i$ (note that  $p_{acc}({i\rightarrow j})=0$ if $j$ is occupied by an obstacle). The general moment-propagation approach to compute the diffusion coefficient is described in \ref{app:MP}.
 
In a homogeneous fluid, successive jumps are uncorrelated and we have:
\begin{equation}\label{eq:D1step}
 D = {1\over 2\Delta t} \bracketed{(\Delta x)^2}
\end{equation}
If the probability to jump to a neighbouring site is equal to one, then $D= a^2/(2d\Delta t)$, where $a$ is the lattice spacing. However, we can reduce the probability that a particle carries out a jump. If we denote the probability that a particle stays on the same site by $1-\alpha$, then
\begin{equation}
 D =\alpha {1\over 2d\Delta t} a^2 \;.
\end{equation}
In systems where the diffusivity varies with position, we can define a factor $\alpha(ij)$ for every link between neighbouring lattice sites $i$ and $j$. It is important to note that if the probability to jump from $i$ to $j$ is reduced, then the probability to jump from $j$ to $i$ should be reduced by the same factor, otherwise the equilibrium distribution over lattice sites would be changed from the Boltzmann distribution.

\subsection{NMR signal and spectrum}

NMR spectroscopy probes the nuclear magnetic response of a sample whose magnetism is perturbed from equilibrium by a radio frequency pulse. After this pulse, the transverse magnetisation of the sample decays. The NMR signal measured during this decay is commonly referred to as the Free Induction Decay (FID). The NMR spectrum is obtained by Fourier transforming the FID signal. In a heterogeneous sample, different nuclei will experience different local magnetic fields and will therefore have different Larmor frequencies. As the nuclei diffuse, their environments, and hence their resonant frequencies, change. The FID signal is the superposition of the signals corresponding to all excited nuclei in the sample. For an ensemble of probed spins, the NMR signal is given by:
\begin{equation}  
G(t)=\langle {\rm e}^{{\rm i}\int_0^t 2\pi\nu_0^i(t').dt'}\rangle
\end{equation}  
where $\nu_0^i(t)$ is the Larmor frequency corresponding to spin $i$ at time $t$, and $\langle ...\rangle$ denotes an average over all spins. The spectrum is then obtained by Fourier transforming this signal following:
\begin{equation}   
\mathcal{F}(k) = \int_{-\infty}^{\infty}dt\; G(t){\rm e}^{-2{\rm i}\pi kt}\;.
\end{equation}   
Using the same approach as for the calculation of the diffusion coefficients (see Appendix~\ref{app:NMR}), we can discretise these expressions and estimate them by the moment-propagation method. 

In practice, it is better to define the resonant frequency $\nu_i$ as the difference beween the Larmor frequency at site $i$, $\nu_0^i$, and the Larmor frequency of the bulk liquid (the reference). The resulting frequencies are typically in the kilo-Hertz range. The timestep $\Delta t$ should be chosen such that $\Delta t\times \nu_{max}\ll 1$, where $\nu_{max}$ is the largest value of $\nu_i$. We note here that the spectral width $SW$, i.e. the range of frequencies that is studied, is given by the dwell time, $dwt$, i.e. the time between two acquisition sampling points, following:
\begin{equation}
SW=\frac{1}{dwt}.
\end{equation}
This spectral width in Hertz can then be converted to a range of chemical shifts in ppm using the Larmor frequency of the studied nucleus ($\nu_0$):
\begin{equation}    
SW(ppm) = \frac{SW (Hz)}{\nu_0}\times 10^6.
\end{equation}    
Note that, the frequency differences (in Hz) between the various resonant frequencies experienced by the spins depend on the applied magnetic field, the frequency increasing linearly with field strength. This is taken into account in the model as will be clear from the study of various magnetic fields in the last section of this article.   

\subsection{Model parameters}

In practical cases, we must map the system that we wish to simulate on a lattice model with discrete lattice sites and timesteps. 
In particular, we must specify:\\
- the distribution of obstacles on the lattice to represent \emph{the porosity of the material}\\
- the distribution of local energy values that account for \emph{the adsorption of ions and molecules to the carbon surface}\\
- the distribution of local Larmor frequencies that account for \emph{the spatial variation of the chemical shifts}.\\
The allocation of these quantities to the different lattice sites will typically be based on structural information or use input data from more microscopic descriptions such as Density Functional Theory or Molecular Dynamics simulations (MD). 

In addition, we must specify the lattice parameter ($a$), i.e. the distance between two lattice nodes in the model, the temperature ($T$) and the timestep ($\Delta t$). These parameters are usually determined by the physical properties of the system to be modelled. The lattice parameter $a$ will usually be defined so that it allows a sufficiently accurate description of the liquid structure. The choice of the timestep is then set by its relation with the diffusion coefficient $D$ and the lattice parameter $a$, namely $D=a^2/(6\Delta t)$ in three dimensional lattices. We then check that the condition $\Delta t\times \nu_{max}\ll 1$ is respected. If the diffusion coefficient or other quantities (free energies for example) correspond to a given temperature, then this temperature should be used in the lattice simulation. 


\section{Application of the lattice method to a two-site exchange system}
\label{Exchange}

\subsection{Parametrisation of a two-site exchange model}

To validate our numerical approach, we show that the model reproduces the lineshape for a two-site exchange model for which analytical solutions exist~\cite{Levitt_book,Cavanagh_book}. 
\begin{figure*}[ht!]
\begin{center}
\includegraphics[scale=0.5]{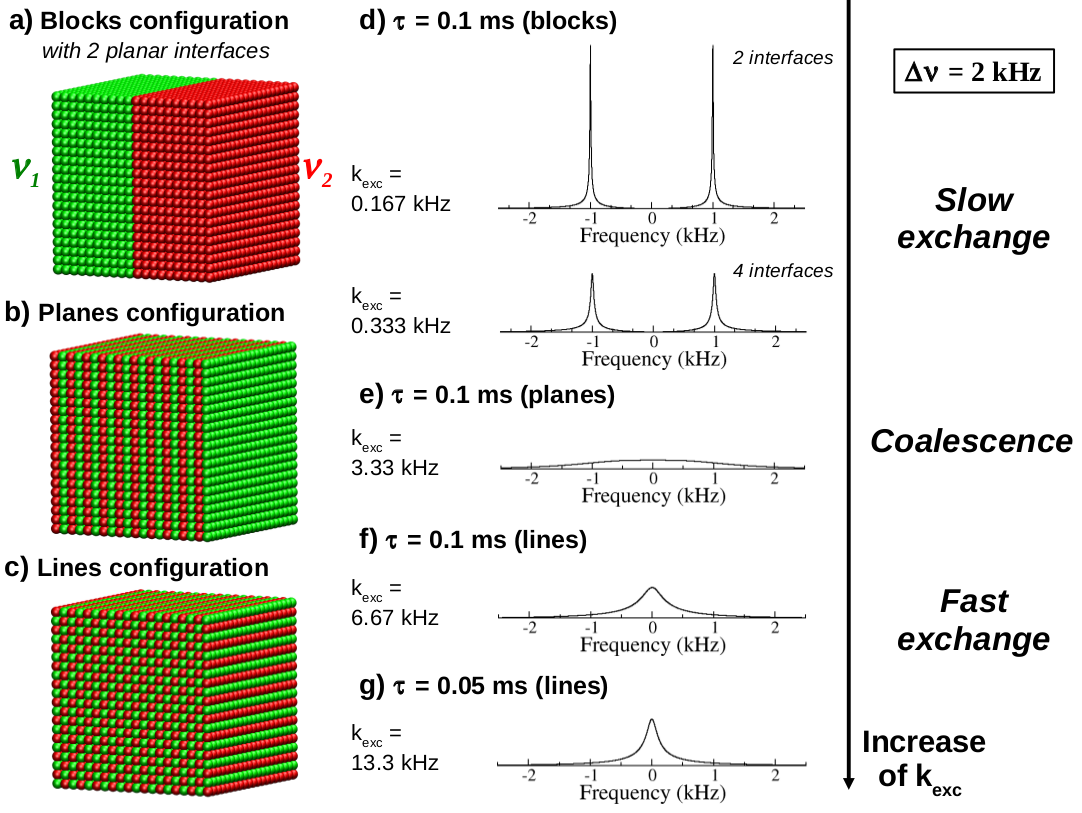}
\end{center}
\caption{Lattice gas model and two site-exchange between sites with frequencies $\nu_1$ (-1~kHz) and $\nu_2$ (1~kHz). Here we illustrate the ability of the model to reproduce coalescence for a simple system. The NMR lineshape typically depends on the ratio between the effective exchange rate $k_{exc}$ between sites with different frequencies and the frequency difference $\Delta \nu=|\nu_1-\nu_2|$. We explore the effect of the spatial distribution of sites with different Larmor frequencies for a given correlation time of 0.1~ms (a,b,c). Here, each sphere represents a lattice site and the color distinguishes sites having a frequency $\nu_1$ or $\nu_2$. The spectra in panel d) illustrate the effect of doubling the number of interfaces, between blocks of distinct frequencies, from two to four: in both cases, two resonances are distinguishable but the resonances are broader when the effective exchange rate increases. e) In the \emph{planes} case, we are close to the coalescence point and we observe a very broad spectrum where we cannot distinguish anymore the two different resonances. f) In the \emph{lines} cases, we note a narrowing of the spectrum which is accentuated in g) by decreasing the correlation time $\tau$.}
\label{exchange}
\end{figure*}
We focus on a simple model in which a spin can move between two different chemical environments, $S_1$ and $S_2$, with equal free energies and equal populations. The process is described by the following equilibrium:
\begin{eqnarray} 
& k & \nonumber \\
S_1 & \rightleftharpoons & S_2\\
& k & \nonumber 
\end{eqnarray} 
where $k$ is the exchange rate between the two sites. The NMR lineshape depends on the ratio between the exchange rate $k$ and the difference in resonant frequencies $\Delta \nu$~=~$|\nu_1-\nu_2|$. Two different regimes can be observed: i) a slow-exchange regime where signals due to the two environments can be distinguished in the spectrum and ii) a fast exchange regime where only a single resonance is observed at a frequency corresponding to an average of $\nu_1$ and $\nu_2$. The limit between these two regimes is known as the \emph{coalescence point} where the two peaks merge into one. If we ignore the spin-spin and spin-lattice relaxation effects, coalescence occurs for an exchange rate equal to:
\begin{equation}  
k_{coal}=\frac{\Delta\omega}{2\sqrt{2}}
\end{equation}  
where $\Delta\omega = 2\pi\Delta\nu = |2\pi\nu_1-2\pi\nu_2|$~\cite{Levitt_book}. 

To investigate two-site exchange with our model, we build a lattice of 40$\times$40$\times$40 sites (64,000 lattice sites in total) where a specific frequency ($\nu_1$~=~-1~kHz or $\nu_2$~=~1~kHz) is assigned to each lattice site. For all the systems investigated, there are no obstacles so all sites are accessible, and the energies on all sites are equal so that all densities are equal. The model is characterised by a correlation time $\tau$ (equivalent to the timestep of the simulation) which is the average time that a diffusing species stays on a site. The inverse of this correlation time thus defines the exchange rate between two sites of the lattice. Not all the jumps between sites will lead to a change of frequency. We thus define an effective exchange rate, $k_{exc}$, which represents the exchange rate for hops between sites with different frequencies, corresponding to the entire system. This is the rate that is measured in the NMR experiments.

In our model, $k_{exc}$ will depend on two key factors: i) the spatial distribution of the sites with different Larmor frequencies on the lattice, i.e. for a given exchange rate, the more contacts there are between sites with different frequencies, the larger $k_{exc}$ will be, ii) the correlation time, i.e. decreasing the correlation time leads to a global increase of the dynamics of the diffusing species and thus an increase of $k_{exc}$. We first explore the effect of the spatial distribution of frequencies by keeping the correlation time $\tau$ constant (equal to 0.1~ms) and arranging chemical environments in different ways. Four different configurations were built (see figure~\ref{exchange}):\\
- \emph{blocks} with 2 planar interfaces: the frequencies of the different sites are distributed in two blocks with equal number of sites, and thus there are only two planar interfaces (one in the middle of the system and one at the edge where periodic boundary conditions are applied) where diffusion will lead to a frequency change;\\
- \emph{blocks} with 4 planar interfaces;\\
- \emph{planes}: the frequencies are distributed such that alternating planes are characterised by $\nu_1$ and $\nu_2$;\\
- \emph{lines}: the frequencies are distributed such that one line in two is characterised by $\nu_1$ and the other by $\nu_2$.\\
 We then investigate the effect of a reduction of the correlation time by a factor two ($\tau$~=~0.05~ms) in the \emph{lines} configurations. We calculate the NMR signals on a time period of 500~ms which corresponds respectively to 5,000 and 10,000 timesteps when $\tau$ equals 0.1~ms and 0.05~ms. This time is longer than the decay time so that there is no truncation of the NMR signal. 

\subsection{From slow exchange to fast exchange}

The NMR spectra predicted for the different setups are given in figure~\ref{exchange}. To relate these results to the outcomes of the analytical models described above, we calculate the effective exchange rate corresponding to the different systems. In all cases, the system is three-dimensional so that a particle will jump in one of the six available directions at each timestep. In the case of the \emph{blocks} configuration with two interfaces, the number of sites where exchange is possible is equal to the number of sites in one plane (40$\times$40) multiplied by two (because exchange can be in both directions) and by the number of interfaces. The average exchange rate will then be given by the probability of being in one of these sites multiplied by the probability of actually jumping in the direction corresponding to a frequency change (1/6):
\begin{equation}       
k_{blocks} = \frac{1}{6}\times\frac{4\times 40\times 40}{64,000}\times \frac{1}{\tau} = 0.167 {\rm~kHz},
\end{equation}
where $\tau$ is equal to 0.1~ms. For the \emph{blocks} configuration with four interfaces, the effective exchange rate doubles to 0.33~kHz. These two frequencies are lower than $k_{coal}$, equal to 4.44~kHz in our case, leading to a slow exchange regime where both peaks are distinguishable. For the case of \emph{planes} and \emph{lines} with a correlation time of 0.1~ms, the number of directions which lead to frequency changes are respectively 2 (out of 6) and 4 (out of 6), giving effective exchange rates of 3.33~kHz and 6.67~kHz, respectively. The first value is very close to the coalescence point and the corresponding spectrum is very broad. The second value falls in the fast exchange regime where peak narrowing is observed. The effective exchange rate is then increased by a factor two using a smaller correlation time of 0.05~ms leading to a further motional narrowing. 

This illustration of a simple two-site exchange model shows that the numerical approach that we propose here can reproduce an albeit simple example of the collapse of the NMR spectrum. But while analytical models can only deal with simple cases such as two-site exchange, the numerical moment-propagation approach can be used to model more complex systems with realistic spatial distributions of resonant frequencies.

\section{Calculation of NMR spectra for an organic electrolyte in a slit pore}
\label{Large}

\subsection{Parametrisation of the lattice model from molecular simulations}

We now focus on the case of slit pores in an attempt to represent what would happen in one class of mesoporous materials. 
As stated in the description of the lattice gas model proposed here, each simulation goes through a parametrisation step where the obstacles' positions, free energies ($E_i$) and frequencies ($\nu_i$) have to be assigned to discrete lattice sites. In the case of a slit pore, the positions of the obstacles are defined by two planes corresponding to the two pore walls. We assume that the slit is perpendicular to the $z$-axis and that there is no variation of free energies or Larmor frequencies with $x$ or $y$. As a consequence of this and the use of periodic boundary conditions, we only need 1~site in the $x$ and $y$ directions. The lattice parameter $a$ is set to 0.5~\r{A} and the number of lattice sites in the $z$ direction, $N_z$, is adapted to the pore width such that $N_z$~=~$(R$/$a)+1$ with $R$ being the pore width. We first focus on a slit pore with a pore width equal to 4~nm, and the number of lattice sites for this system is thus 1$\times$1$\times$81. 

\begin{figure}[ht!]
\begin{center}
\includegraphics[scale=0.6]{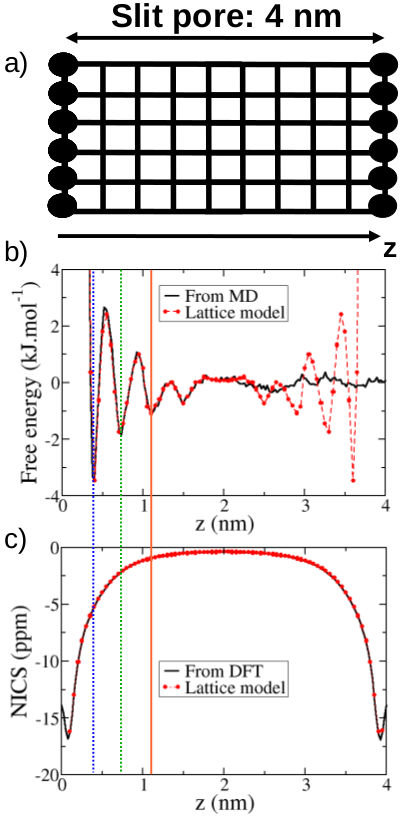}
\end{center}
\caption{Parametrisation of the lattice model from molecular simulations for a 4~nm slit pore. a) For planar surfaces, two planes of obstacles are set at the extremities of the lattice in the $z$ direction, particles moving freely in $x$ and $y$ directions. b) The free energies on the accessible sites are mapped onto the free energy profiles obtained for the BF$_4^-$ anions from MD simulations performed on [BMI][BF$_4$] dissolved in acetonitrile (1.5~M)~\cite{Merlet13b}. The vertical lines on the figure indicate the positions of the different adsorbed layers corresponding to free energy minima. c) The NICS are parameterised with DFT calculations following the same mapping procedure as for free energies. Here, the NICS correspond to the case of nuclei located on a line joining the center of masses of two parallel circumcoronene molecules~\cite{Forse14}.} 
\label{param}
\end{figure}

The values of the site free energies $E_i$ were fixed using existing results obtained from MD simulations. Specifically, we use the simulation results for butylmethylimidazolium te\-tra\-fluoroborate dissolved in acetonitrile ([BMI][BF$_4$] in ACN, 1.5~M) at the interface with a graphite surface. As described in ref.~\cite{Merlet13b}, it is possible to calculate densities and consequently free energy profiles from MD simulations. Here, we use the free energy profiles for the BF$_4^-$ anion. At the interface with a planar surface, ions tend to organise themselves in a layered structure which can extend up to several molecular diameters away from the surface. The free energy profiles are thus characterised by peaks at various distances from the interface as can be seen in figure~\ref{param}. We limit ourselves to $^{11}$B NMR but note that we would expect qualitatively similar results for $^{19}$F and $^1$H NMR.
 
To assign free energies to the lattice sites, we simply map the MD values on the different sites, i.e. for each lattice site, we calculate its distance to the pore wall and give it the corresponding energy value from the realistic free energy profile obtained through MD simulations~\cite{Merlet13b} (figure~\ref{param}). There are three comments we can make at this point: i) the accuracy of the results obtained through the lattice method will depend on the accuracy of the initial MD simulations, ii) the accuracy of the final results will also depend on the lattice parameter chosen, i.e., the finer the grid, the better the representation, and iii) while we do not explicitly include correlated motions of the ions in the model, the correlation effects that result in the layered structure of the liquid at the interface are accounted for by our model. To investigate the first point, we also calculate the NMR spectra for a flat profile, where the free energy is constant across the pore. In this case, we chose a closest distance of approach of 0.32~nm, in agreement with the MD results~\cite{Merlet13b}. 

The frequencies, which correspond to the Larmor frequencies of the nuclei in different local environments for a real system, follow from the DFT calculations of ref.~\cite{Forse14}. The frequencies on the lattice sites are simply mapped using the DFT results as was done for the free energies (see figure~\ref{param}). By doing this, we make a number of assumptions. Firstly, we use the chemical shifts calculated with the Gaussian software~\cite{Juselius99} with the Nucleus Independent Chemical Shift (NICS) approach, i.e. we assume that the chemical shifts originate from ring current effects and that the nature of the ion and its charge do not influence this quantity. Secondly, the calculations reported were not performed on graphene~\cite{Forse14}, which is difficult to treat from an ab-initio point of view, but rather on different aromatic molecules such as coronene, circumcoronene and dicircumcoronene. The molecular size has an impact on the NICS values calculated and here, we will show results for these three aromatic molecules. We note here that the chemical shift also depends on the lateral position of the ions over the surface~\cite{Forse14,Xing14}. While this is not investigated here, this spatial dependence is also expected to impact the NMR results.

\begin{figure*}[ht!]
\begin{center}
\includegraphics[scale=0.5]{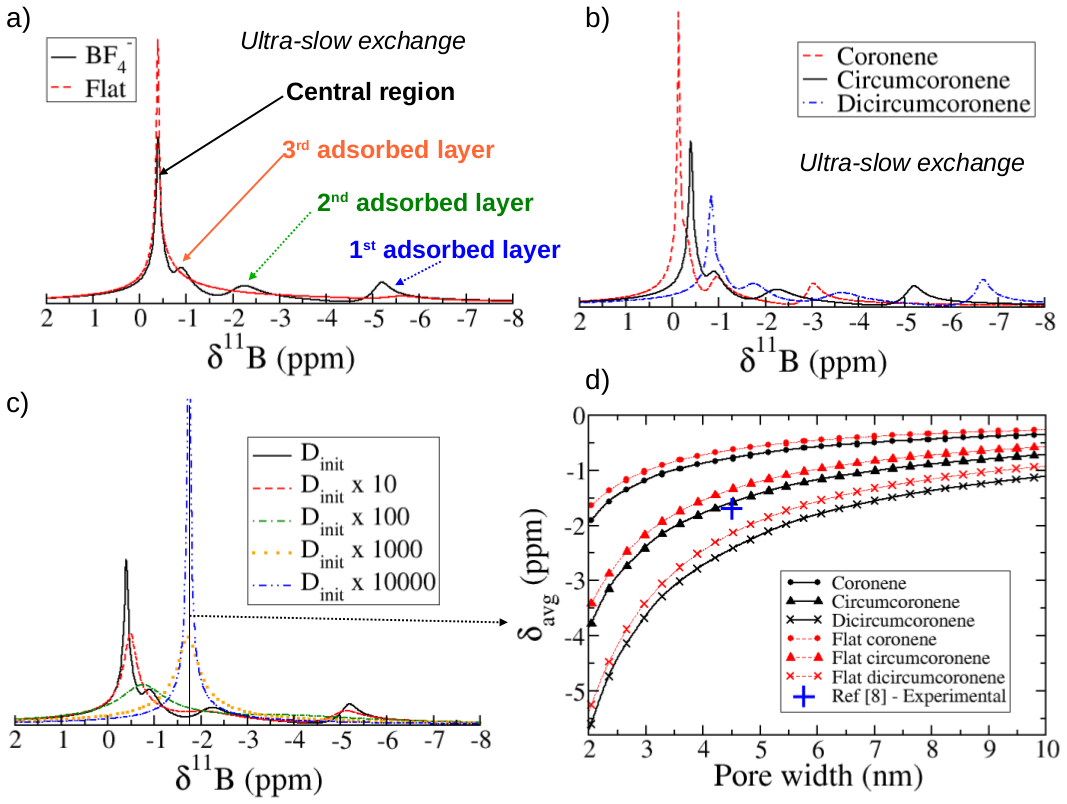}
\end{center}
\caption{NMR spectra for the case of a slit mesopore (4~nm) with an organic electrolyte ([BMI][BF$_4$] in acetonitrile) and variation of the average chemical shift with pore width. a) NMR spectra for ultra-slow anion diffusion. Two very different spectra are obtained if we assume either a realistic free energy profile extracted from molecular simulations~\cite{Merlet13b} (solid black line) or a flat energy profile in which the anionic density is constant across the pore (dashed red line). b) NMR spectra calculated with different NICS profiles. The size of the aromatic molecule used in the DFT calculation has a large impact on the calculated spectrum: the larger the molecule, i.e. the graphitic carbon domain, the more shifted the peaks are. c) The effect of slow/fast exchange is investigated through a variation of the diffusion coefficient from $D_{init}$~=~84.7~10$^{-19}$~m$^2$.s$^{-1}$ to $D$~=~84.7~10$^{-15}$~m$^2$.s$^{-1}$. A faster exchange leads to the broadening and merging of the peaks. d) Variation of the average chemical shift as a function of pore width for various parametrisations with different NICS and free energy profiles. The lattice model associated with the circumcoronene DFT calculations is in good agreement with a single experimental point corresponding to a CMK-3 carbon material with an average pore size of 4.5~nm~\cite{Borchardt13}.}
\label{spectra}
\end{figure*}

Finally, we need to determine the timestep $\Delta t$ corresponding to the lattice parameter $a$ chosen (0.5~\r{A}). Molecular dynamics simulations of the bulk solution of [BMI][BF$_4$] in acetonitrile have reported a diffusion coefficient of 84.7~10$^{-11}$~m$^2$.s$^{-1}$ for the BF$_4^-$ anion~\cite{These_Merlet}. The timestep would thus be equal to 4.92~10$^{-13}$~s following $D= a^2/(2d\Delta t)$. We would like to stress here that if we were to simulate a complete NMR spectrum with the parameters used experimentally, it would be very computationally costly. Indeed, the experimental dwell time (the time between two FID points) is chosen depending on the desired spectral width and is typically of the order of 5~$\mu$s so we would have to perform a simulation where each data point from the experimental NMR signal corresponds to more than one million timesteps in our model. Nevertheless, we will show that this is not a problem here as the lattice model allows us to investigate slower dynamics, which provides information that allows to understand experimental spectra. In our model, the diffusion coefficient is increased by changing the timestep while keeping all other parameters constant. We note that the simulation time is kept constant for all the simulations, so when the timestep is decreased by a factor $K$, the total number of steps is also increased by a factor $K$ and the points are sampled to perform the Fourier transformation on the same number of values. All the simulations are long enough to observe the entire decay of the NMR signal.      

\subsection{NMR spectra for ultra-slow diffusion and effect of increasing the ion dynamics}

We first calculate NMR spectra for $^{11}$B in a 4~nm slit pore for the hypothetical case of ultra-slow diffusion ($D_{\rm init}$~=~84.7~10$^{-19}$~m$^2$.s$^{-1}$, which corresponds to a timestep of $\Delta t$~=~4.92~10$^{-5}$~s; a~=~0.5~\r{A}). The NMR spectra obtained with the lattice model using the realistic free energy profile from MD simulations~\cite{Merlet13b} and the flat free energy profile are shown in figure~\ref{spectra}. It is clear that assuming a flat energy profile leads to large differences in the resulting NMR spectrum and the disappearance of many features. In the case of the realistic free energy profile, we can identify different peaks corresponding to the different environments experienced by the anions which tend to organise themselves in layers, as can be seen from the free energy profile (see figure~\ref{param}). The first adsorbed layer gives rise to the peak at most negative chemical shift while ions in the second and third are observed at smaller shifts. The sharp peak close to 0~ppm corresponds to the central region of the slit pore, which is only weakly affected by the graphitic carbon surfaces. In the case of a flat energy profile, we observe a single resonance, with a shoulder on the negative frequency side.

We now investigate the impact of the NICS profiles on the calculated spectra. We focus on three different molecules, namely coronene, circumcoronene and dicircumcoronene which have molecular areas respectively close to 0.362~nm$^2$, 0.981~nm$^2$ and 1.911~nm$^2$. The DFT calculations on these molecules show that the larger the molecular size, the larger the chemical shifts, due to the increased number of rings contributing to ring current effects~\cite{Forse14}. The NMR spectra obtained with the lattice model using the different NICS profiles (figure~\ref{spectra}) are qualitatively similar but mirror the effect of the DFT calculations in the sense that larger molecules lead to NMR spectra shifted to lower frequencies. 
 
The effect of increasing the diffusion coefficient of the anions on the resulting NMR spectra is explored by working with a realistic free energy profile for the anions and the NICS profile corresponding to circumcoronene. We start with a diffusion coefficient $D_{init}$~=~84.7~10$^{-19}$~m$^2$.s$^{-1}$ and increase it by several orders of magnitude to reach $D$~=~10,000$\times D_{init}$~=~84.7~10$^{-15}$~m$^2$.s$^{-1}$ (figure~\ref{spectra}). When the diffusion is increased by a factor 10 or 100, the peaks corresponding to the different environments broaden and start to merge. With an increase of a factor 1,000, only one peak appears in the spectrum and it becomes sharper as the diffusion is increased further. As the real diffusion coefficient is much larger than the ones studied here (by a factor of 10$^4$), we should expect that a single very sharp peak would be observed experimentally for a porous carbon consisting of slit pores with a monodisperse pore size distribution: on the timescale probed by the NMR experiment, for the organic electrolyte investigated here, the anions would explore the entire slit pore and only an average chemical shift would be observed. 

A much broader lineshape is observed experimentally than the one predicted by the simple slit-pore model~\cite{Forse13,Forse14,Borchardt13}, and the experimental lineshape depends significantly on the temperature~\cite{Forse14b}. The current results demonstrate that the experimentally observed lineshape is not related to diffusion in a pore but to the fact that ions explore a distribution of pore sizes. Indeed, the experimental carbons usually show some disorder with a distribution of pore sizes and pore geometries. A realistic model would thus have to combine information about the average shift within pores of different sizes and the distribution of exchange rates between pores. With this information, an even more coarse-grained lattice model can be constructed in which each accessible lattice site is a pore. The average chemical shifts corresponding to the various pore sizes can be determined using the current approach and used as an input in this new system. 

\subsection{Variation of the average chemical shift as a function of pore width}

In this part, we investigate the variation of the average chemical shift as a function of pore width with the following aims: to i) obtain insight in the global trend of the variation and the impact of the input parameters (free energy and NICS profiles) on the resulting data and ii) produce the information necessary for the parametrisation of a new model representing a carbon particle with a realistic pore size distribution. We will thus calculate the average chemical shift for a range of pore widths ranging from 2 to 10~nm for different setups. 

We note here that, for the determination of the average chemical shift, we do not need a lattice model calculation. The average chemical shift is simply given by the integrated NICS profile weighted by the free energy profile:
\begin{equation}
\delta_{avg}=\int_0^R  \frac{\exp \left ( \frac{-E(r)}{k_B T}\right ) }{\int_0^R \exp \left ( \frac{-E(r')}{k_B T}\right )dr' }\times \nu(r)dr.
\end{equation}
This approach was suggested in a previous study by Xing~\emph{et al.}~\cite{Xing14} where the authors apply this methodology to obtain insights into $^1$H NMR spectra of water in porous carbon materials. While they consider only a flat energy profile for the hydrogen atoms, we show here that the inclusion of a realistic free energy profile leads to some variations in the obtained average chemical shifts.

The plots showing the variation of the average chemical shift as a function of pore width for the various conditions investigated are given in figure~\ref{spectra}. All the curves are qualitatively similar showing that the absolute value of the chemical shift increases exponentially with decreasing pore width. As a general result, the graphitic domain size has a larger influence on the results than the use of a realistic or a flat free energy profile. Nevertheless, the influence of the free energy profile should not be neglected as it can lead to relative errors of up to 25~\% for the flat profile compared to the realistic profile. We note that the use of a flat energy profile, corresponding to a constant density across the pore, leads to an overestimation of the average chemical shifts, i.e. the values are shifted to larger frequencies compared to the realistic free energy profile values. This is a consequence of the neglect of high anionic density close to the surface that exists in the layered structure.

Figure~\ref{spectra} also shows an experimental result obtained by Borchardt~\emph{et al.}~\cite{Borchardt13} for a solution of tetraethylammonium tetrafluoroborate ([TEA][BF$_4$]) adsorbed in an ordered mesoporous carbon (CMK-3) with an average pore size of 4.5~nm: the $^{11}$B NMR spectrum for adsorbed anions shows a shift of -1.7~ppm compared to the free electrolyte. This (albeit single) experimental measurement provides a good point of comparison with our model for several reasons. Firstly, the liquid investigated experimentally is a 1~M solution of [TEA][BF$_4$] in ACN which is close to the present system consisting of a 1.5~M solution of [BMI][BF$_4$] in ACN~\cite{Merlet13b}. Secondly, for mesopores, the pore walls should appear microscopically as being close to planar surfaces so that the approximation of pores by slit pores might be acceptable. 
We see on figure~\ref{spectra} that this experimental result seems to be in good agreement with the trend obtained with the lattice model and the circumcoronene molecule. While this agreement is partly fortuitous as there is no obvious reason at this point suggesting that the circumcoronene molecule should be a better model than the other aromatic molecules, this is very promising for further applications of the lattice model. While Borchardt~\emph{et al.}~\cite{Borchardt13} and other authors~\cite{Wang13,Forse13} report NMR results for other carbon materials with average pore sizes in the range of 1~nm, these were not investigated here as the current model for slit pores is parameterised through MD simulations of liquid inside a mesopore~\cite{Merlet13b}. 

\section{NMR spectra calculation for a porous carbon particle}

\subsection{Parametrisation of the lattice model for a carbon particle}

In the final part of this article, we use an even more coarse-grained lattice model to represent a carbon particle with a realistic pore size distribution and the exchange of ions between pores of different sizes. In this new setup, each lattice site represents an entire pore, with a given pore size. We apply this model to study how the spatial distribution of pore sizes in a carbon particle affects the resulting NMR spectra. In what follows, we will consider a model that resemble the experimental system of Borchardt~\emph{et al.}~\cite{Borchardt13}. As before, we have to assign a number of quantities to each lattice site before performing the simulations, namely in this model, i) the pore size, ii) the free energy, iii) the chemical shift and iv) the energy barriers associated with the exchange between each lattice site and its neighbours.

\begin{figure*}[ht!]
\begin{center}
\includegraphics[scale=0.6]{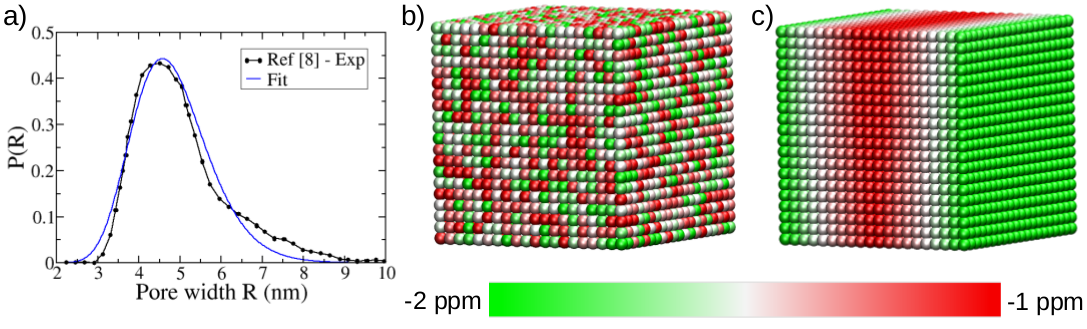}
\end{center}
\caption{Pore size distribution and spatial distribution of chemical shifts. a) Pore size distribution used in the model, obtained by fitting experimental data corresponding to a CMK-3 carbon~\cite{Borchardt13} with a log normal distribution. b) and c) Color representation of the spatial distribution of shifts in the lattice: each sphere represents a lattice site. The shifts can be distributed randomly (b) or following a gradient in one of the three dimensions (c).}
\label{psd}
\end{figure*}

To parametrise the model, we first need to choose a pore size distribution. In their work, Borchardt~\emph{et al.}~\cite{Borchardt13} report an experimental pore size distribution obtained through nitrogen adsorption. The reported experimental pore size distribution presents two maxima: one close to 1~nm and the other close to 4.5~nm. As most of the pore volume in this material corresponds to pores larger than 2~nm (93~\% of the total pore volume), we only consider this part of the curve in our model. We fit a log normal pore size distribution to the experimental curve. The mean of this log normal distribution is 1.558 while the standard deviation is 0.193. As shown in figure~\ref{psd}, the fit is quite good and the log normal distribution seems to be a good choice to represent the experimental pore size distribution. 

The free energy associated to each site depends on the amount of liquid adsorbed at this site and hence  on the pore width ($R$), the pore surface ($S$), and the free energy profile inside the pore. As before, we will assume that all the pores are slit pores. For each lattice site, we choose randomly a pore width following the log normal distribution described above. We then need to choose a pore surface to completely define or encapsulate the pore volume associated with the lattice site considered. We choose to work with a distribution of pore surfaces centered around the area of a circumcoronene molecule, since the experimental result~\cite{Borchardt13} was close to the simulation results for this molecule, and intermediate between the coronene and dicircumcoronene areas for which we also have chemical shifts values. The distribution is arbitrarily chosen to be log normal with a mean of -0.1 and a standard deviation of 0.25. Once the pore width and the pore surface have been determined, the free energy of the lattice site is defined so that the density inside the pore, $\rho_i$, is proportional to the integrated density inside the slit pore multiplied by the pore surface:
\begin{equation}
\rho_i \propto \int_0^R {\rm e}^{-\beta E(r)}dr \times S.        
\end{equation}
In this way, we take into account both the volume of the pore and the liquid structure inside it to define the free energy of the lattice sites.

The next step is to assign a chemical shift to each lattice site. This is again done according to the pore width and pore surface previously chosen. The value of the chemical shift for a lattice site is intrapolated from the values calculated for the three aromatic molecules. For example, for a pore width $R$, and a pore surface $S$, intermediate between the coronene area and the circumcoronene area, the chemical shift is chosen as a weighted average of the coronene value at $R$, $\delta_{coron}(R)$ and of the circumcoronene value at $R$, $\delta_{circum}(R)$:
\begin{equation} 
\delta_i = \delta_{coron}(R)+\frac{\delta_{circum}(R)-\delta_{coron}(R)}{S_{circum}-S_{coron}}\times (S-S_{coron})
\end{equation} 
where $S_{coron}$, $S_{circum}$ and $S_{dicircum}$ are taken to be respectively 0.362~nm$^2$, 0.981~nm$^2$ and 1.911~nm$^2$. 

In our lattice model representing a carbon particle, it is not particularly relevant to set a lattice parameter as we do not have a clear idea of what would be a realistic distance between pores of different sizes and we do not know how the diffusion coefficients are affected by the confinement. In contrast, it is essential to be able to vary the exchange rate between pores, as this rate will affect
the temperature dependence  of  the NMR lineshape.  We will assume that the presence of (positive) energy barriers $Ea(ij)$ reduce the forward and backward jump rate between sites $i$ and $j$.  
Compared to free diffusion, the probability of jumping from a site $i$ to a site $j$ is then reduced by a factor $\alpha(ij)$ equal to:
\begin{equation}
\alpha(ij)=\exp \left ( \frac{-Ea(ij)}{k_BT} \right).
\end{equation}       
Note that, as mentioned in section~\ref{Method}, the probability of jumping from $j$ to $i$ has to be reduced by the same factor $\alpha(ij)$ in order to maintain detailed balance. We do not know the inter-pore jump rates {\em a priori}.  In what follows, we will assume that the energy barriers obey a Gaussian distribution with a mean value $E_{mean}$ and a standard deviation equal to 0.2, irrespective of their position in the lattice.  

Following this parametrisation scheme, we build three different models of carbon particles with the same pore size distribution. We use cubic lattices with a dimension of 20$\times$20$\times$20. To check a possible size dependence of the resulting NMR spectra, we also performed the calculations at room temperature with lattices of dimensions 30$\times$30$\times$30 and 40$\times$40$\times$40. The obtained spectra are very similar for the different lattice sizes so that we consider only the smaller lattice in the rest of this article. All calculations are performed with a time step equal to 5~$\mu$s and the simulations are run for 50,000 time steps (0.25~s), which is  much longer than the decay time of the NMR signal. 

We first build a model where the various pore sizes, and consequently the chemical shifts associated, are distributed randomly over the lattice. This situation is represented in figure~\ref{psd}b). For this model, we choose the mean value of the energy barriers $E_{mean}$ such that the calculated linewidth (full width at half maximum peak intensity) at room temperature is equal to the experimental value of 0.45~ppm obtained by Borchardt~\emph{et al.}~\cite{Borchardt13}. This results in a value for $E_{mean}$ equal to 11.8~kJ.mol$^{-1}$. Note that the calculation is performed with a spectrometer frequency of 300~MHz as the experiments were done with this condition. This first model will be referred to as \emph{Random-11.8}. To investigate the effect of the spatial distribution of shifts, we then create a setup where the pore sizes are organised in an ordered way such that there is a gradient of chemical shifts in one direction of the lattice as illustrated in figure~\ref{psd}c). The second model, referred to as \emph{Grad-11.8}, is a lattice where the chemical shifts are ordered following a gradient and the mean energy barrier is $E_{mean}$~=~11.8~kJ.mol$^{-1}$. This model allows us to probe the effect of the gradient on the NMR lineshape while keeping all other parameters constant. Finally, the third model, referred to as \emph{Grad-1.7} is a lattice where the chemical shifts are ordered following a gradient and $E_{mean}$ is set equal to 1.7~kJ.mol$^{-1}$ to recover a linewidth of 0.45~ppm at room temperature. The comparison between \emph{Random-11.8} and \emph{Grad-1.7} allows us to probe the different behaviours of the random and gradient spatial distributions for two systems giving the same linewidth at room temperature.           

\subsection{Investigation of various parameters on the NMR spectra of the model carbon particles}

\begin{figure*}[ht!]
\begin{center}
\includegraphics[scale=0.6]{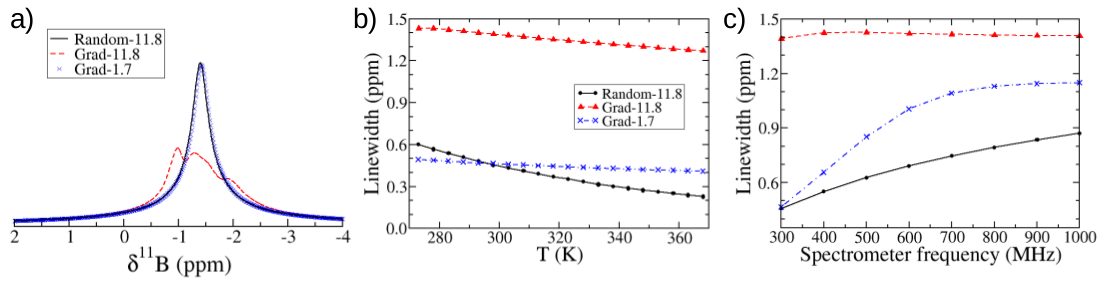}
\end{center}
\caption{NMR spectra and linewidths for different setups. a) NMR spectra obtained at room temperature for random/gradient distributions of shifts and barrier heights distributions centered around $E_{mean}$~=~11.8 or 1.7~kJ.mol$^{-1}$. The gradient of shifts (dashed red line) leads to a much broader spectrum than the random distribution (black solid line). In the gradient case, when the barrier heights distribution is changed to reproduce the experimental linewidth at room temperature (45~ppm~\cite{Borchardt13}), the spectrum (blue crosses) is superimposed with the one calculated for the random case. b) Linewidth as a function of temperature for the different setups. c) Linewidth as a function of spectrometer frequency.}
\label{randgrad}
\end{figure*}

We first calculate the NMR spectra corresponding to the three models \emph{Random-11.8}, \emph{Grad-11.8} and \emph{Grad-1.7} at room temperature (\ref{randgrad}). Comparing  the results for  \emph{Random-11.8} and \emph{Grad-11.8}, which differ only by the spatial distribution of the shifts, we note that they yield very different NMR spectra. The spectrum is much broader in the case of the gradient than in the case of the random distribution. This is not surprising because, in the gradient model it takes longer to sample all Larmor frequencies than in the random model, where a single jump can dramatically change the Larmor frequency. The NMR spectrum for \emph{Grad-11.8} mirrors the asymmetrical distribution of chemical shifts resulting from the log normal distribution of pore sizes. By decreasing the energy barrier from 11.8 to 1.7~kJ.mol$^{-1}$ in the gradient configuration, we recover the NMR spectrum observed for \emph{Random-11.8}.   

The difference in energy barrier for \emph{Random-11.8} and \emph{Grad-1.7} leads to different conclusions concerning the exchange rates and diffusion times in the pores. The relatively long exchange times necessary to observe lineshape variations on the experimental time scales could originate from different factors: a slow diffusion in the carbon particle due to confinement or a slow exchange between pores due to either small interpore connections or large distances between pores. From the simulations, we can estimate the exchange rate between pores needed to reproduce the experimental linewidth of 0.45~ppm~\cite{Borchardt13}. For \emph{Random-11.8}, the exchange rate is equal to 1.75~kHz and corresponds to a correlation time of 570~$\mu$s. For \emph{Grad-1.7}, the exchange rate is much higher, equal to 101~kHz, and corresponds to a correlation time of 9.9~$\mu$s. If we assume that the diffusion coefficient in the porosity is the same as the bulk diffusion coefficient~\cite{These_Merlet}, i.e. equal to 84.7~10$^{-11}$~m$^2$.s$^{-1}$, and that the long exchange times are due to large distances between pores, we can estimate the distance between pores. In the case of \emph{Random-11.8} and \emph{Grad-1.7}, we obtain distances between pores of different sizes respectively equal to 1,702~nm and 224~nm. These values seem very high considering that the particle sizes are usually in the micrometer range. If we assume that the diffusion coefficient is two orders of magnitude smaller in the particle compared to the bulk, we estimate distances between pores of around 170~nm for \emph{Random-11.8} and 22.4~nm for \emph{Grad-1.7}. While the distance still seems quite large for the random distribution as the average pore size, 4.5~nm, is much smaller than 170~nm, the estimation for \emph{Grad-1.7} seems realistic. The point here is that, while we cannot draw unambiguous conclusions about the origin of the slow exchange, we can see that the estimations are strongly affected by the spatial distribution of shifts. 

We now explore the effect of temperature applied magnetic field strength on the NMR lineshapes, to determine if further information on porosity can be extracted (figure~\ref{randgrad}). Different magnetic fields are usually labelled according to the Larmor frequency of $^1$H in that field: we will refer to this quantity as the spectrometer frequency. We first focus on the temperature effect. For all the systems, when the temperature increases, the diffusion of the ions, or equivalently the exchange between different pores, increases leading to a sharpening of the spectra and thus a decrease of the linewidth. Nevertheless, the temperature dependence is not the same for the different systems. The two gradient cases show a slower decrease of the linewidth with temperature compared to the random case. Moreover, the gradient curves appear to be almost linear over this temperature range while for the random case the temperature dependence of the linewidth appears roughly exponential.  Over a wider temperature range, all curves should show an exponential behaviour as a result of the exponential dependence of the transition probability on temperature. The interesting point here is that, while the \emph{Random-11.8} and \emph{Grad-1.7} have the same linewidth at room temperature, this linewidth does not vary in the same way for the two systems. 

The last parameter that we explore is the applied magnetic field by varying the spectrometer frequency from 300~MHz to 1~GHz. The results show that the dependence of the linewidth on the spectrometer frequency is strikingly different for the three systems. The linewidth is almost constant for \emph{Grad-11.8}, showing that the distribution of shifts in the spectrum corresponds to the actual distribution of shifts in the system. In constrast, for \emph{Random-11.8} and \emph{Grad-1.7}, the linewidth depends dramatically on the applied magnetic field. 

Although not investigated here, we can expect that very different ions could lead to shifts in the NMR spectra as the adsorption profiles will not be the same and larger molecules would not be able to access all the pores accessible to small molecules. In conclusion, analysing a range of NMR experiments using the lattice model presented here, should provide a promising tool for characterising porous carbon structures.   

\section{Conclusion}

In this work, we have described a coarse-grained simulation method to predict diffusion coefficients and NMR spectra of probe particles diffusing in a porous carbon matrix. The model that we propose allows us to account for relevant microscopic information whilst exploiting the computational efficiency of the `moment-propagation' approach, a method that allows us to account for {\em all} possible trajectories that particles could follow in a discretised model of a porous network. 

Our simulation method is able to reproduce the coalescence effect observed experimentally, the results agree with analytical solutions in the case of a two-site exchange model. The model allows NMR spectra for much more complex systems to be predicted as it can deal with any set of frequencies and spatial distributions of these frequencies. In particular, we show that the model can be parametrised realistically using input from molecular simulations such as free energy profiles, to represent the molecular/ionic adsorption, and nucleus independent chemical shifts estimated through DFT calculations, to account for the spatial dependence of the chemical shifts of the anions within the carbon pores. 

Parametrisation was performed for an organic electrolyte confined in slit mesopores of various pore sizes and we could show that, while a number of environments would be observed if the diffusion of probed species was ultra-slow, the exchange rates involved in experiments lead to the detection of a single resonance. This peak is observed for an average chemical shift which depends both on the pore size and on the adsorption profile of the studied species. In this work, we also investigate the effect of the molecular size chosen for the chemical shifts calculations on the resulting NMR spectra and show that the graphitic domain size can lead to large variations of the average chemical shifts. This can be seen as a challenge from the parametrisation point of view but it can also be seen as an opportunity to get insights into these domain sizes from NMR while it is not probed from other characterisation techniques.

The last part of this article focused on the parametrisation of an even more coarse-grained  lattice model that can be used to represent a carbon particle with a realistic pore size distribution. In this model, each lattice site represents a pore with a given pore size. The model allows us to explore various spatial distributions of the pore sizes and various conditions, such as applying different temperatures and magnetic fields,  which can be related to experimental conditions. While some of the parameters in this model are known from microscopic simulations, others can be estimated by comparing computed and experimental spectra for a range of temperatures and magnetic fields. Such a comparison yields novel insights into the structure of porous carbon materials and the structure of the liquid inside the pores. In particular, this new lattice model is expected to provide new insights into \emph{in situ} NMR experiments performed on supercapacitors. Moreover, because of its versatility, the lattice model is a powerful tool to investigate a full range of materials, for which NMR parameters can be determined, including battery and fuel cells materials.  

\section*{Acknowledgements}

C.M. acknowledges the School of the Physical Sciences of the University of Cambridge for funding through an Oppenheimer Research Fellowship. C.M., A.C.F., J.M.G. and C.P.G. acknowledge the Sims Scholarship (A.C.F.), EPSRC (via the Supergen consortium, J.M.G.), and the EU ERC (via an Advanced Fellowship to C.P.G.) for funding. A.C.F. and J.M.G. thank the NanoDTC Cambridge for travel funding. D.F. acknowledges EPSRC Grant No. EP/I000844/1.

\appendix
\section{Discrete Green-Kubo relation}
\subsection{Self-diffusion coefficient}\label{app:DIFF} 
We use the Einstein expression relating diffusion and mean-square displacement $\bracketed{\Delta x^2 (t)}$
\begin{equation}
\bracketed{\Delta x^2 (t)} =2Dt
\end{equation}
where $D$ is the self diffusion coefficient of the particles and $t$ is the time.
If particles move by a sequence of jumps $\Delta x_i$ of magnitude $a$, then
\begin{equation}\label{eq:einstein1}
\bracketed{\Delta x^2 (t)} =\bracketed{\left(\sum_{i=1}^N \Delta x_i\right)^2}
\end{equation}
where $N$ is the number of jumps such that $N\Delta t=t$.
We can rewrite Eqn.~\ref{eq:einstein1} as
\begin{equation}\label{eq:einstein2}
\bracketed{\Delta x^2 (t)} =\bracketed{\sum_{i,j=1}^N \Delta x_i\Delta x_j}
\end{equation}
It is convenient to separate the terms with $i=j$ and $i\ne j$:
\begin{equation}\label{eq:einstein3}
\bracketed{\Delta x^2 (t)} =\bracketed{\sum_{i=1}^N (\Delta x_i)^2} +\bracketed{\sum_{i\ne j}^N \Delta x_i\Delta x_j}
\end{equation}
In the second term on the right hand side the terms $i,j$ and $j,i$ contribute equally and hence
\begin{equation}\label{eq:einstein4}
\bracketed{\Delta x^2 (t)} =\bracketed{\sum_{i=1}^N (\Delta x_i)^2} +2\bracketed{\sum_{i< j}^N \Delta x_i\Delta x_j}
\end{equation}
Hence
\begin{equation}\label{eq:GK1}
Dt = D N\Delta t= {1\over 2} \sum_{i=1}^N \bracketed{(\Delta x_i)^2} +\sum_{i< j}^N \bracketed{\Delta x_i\Delta x_j}
\end{equation}
If $N\delta t$ is much larger than the decay time of  $\bracketed{\Delta x_i\Delta x_j}$, we can write:
\begin{equation}\label{eq:GK2}
D = {1\over 2\Delta t} \bracketed{(\Delta x)^2} +{1\over \Delta t}\sum_{j=2}^\infty \bracketed{\Delta x_1\Delta x_j}
\end{equation}
This expression is the discrete analog of the Green-Kubo relation
\begin{equation}\label{eq:GK3}
D = \int_0^\infty dt\; \bracketed{v_x(0)v_x(t)}
\end{equation}
To emphasise this analogy we interpret $\Delta x_i/\Delta t$ as the `velocity' at timestep $i$. Then we obtain:
 \begin{equation}\label{eq:GK4}
D = {1\over 2} \bracketed{v_x^2}\Delta t +\sum_{j=2}^\infty \bracketed{v_x(1)v_x(j)}\Delta t
\end{equation}
Note, however, that we do not assume that the above expression is valid in the diffusive regime. In particular, for diffusion in a homogeneous medium, we have (for not too short $\Delta t$, so that $ \bracketed{v_x(1)v_x(j)}=0$):
\begin{equation}
{1\over 2}\bracketed{v_x^2}\Delta t = {1\over 2}a^2/\Delta t = D \Delta t/\Delta t = D 
\end{equation}
\subsection{Moment-propagation approach}\label{app:MP}
The expression for the diffusion constant given in Eqn.~\ref{eq:GK2} suggests that we should average  Eqn.~\ref{eq:GK2} over all trajectories that a particle can follow. As the number of trajectories of a walk of length $N$ is equal to $z^N$, where $z$ is the lattice coordination number, it would seem that brute-force averaging over all trajectories is not feasible, yet that is exactly what can be achieved using a recursive numerical scheme (`moment propagation'). 
\setcounter{figure}{0}
\counterwithin{figure}{section}
\begin{figure*}[ht!]
\begin{center}
\includegraphics[scale=0.5]{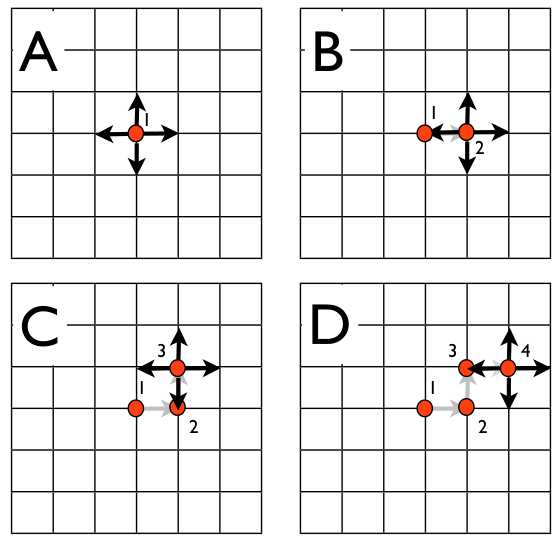}
\end{center}
\caption{Consider a particle starting a random walk at site $i=1$ (A). The probability that this particle is located at a site $i$ is given by the normalised Boltzmann weight $p_B(i)=c e^{-\beta U(i)}$.
In one timestep, the particle can move with a probability ${1\over z} p_{acc}(1\to j)$ to any of the neighbouring sites $j$. For simplicity, we focus our attention on the step to site $j=2$, but our arguments apply to all $z$ neighbours of $1$. 
The displacement vector for the step from site $1$ to site $j=2$ is $\Delta {\bf x}_{12}$. To correlate $\Delta {\bf x}_{12}$ with the subsequent displacement vector starting at $2$, we have to compute the average   $\bracketed{\Delta {\bf x}_2}\equiv \sum_{k=1}^z {1\over z} p_{acc}(2\to k)\Delta {\bf x}_{2k}$ (i.e. the weighted sum over the dark arrows shown in (B)). To get the complete contribution to the correlation function, we have to average the contributions of all $z$ sites neighbouring site $1$. To obtain $ \bracketed{\Delta {\bf x}_1\cdot\Delta {\bf x}_3}$, we have to repeat the same procedure for all sites that can be reached in two steps from site $1$, e.g. site $3$ (see (C)). Similarly, to obtain $ \bracketed{\Delta {\bf x}_1\cdot\Delta {\bf x}_4}$, we must consider all sites that can be reached in three steps from site $1$, e.g. site $4$.}
\label{fig:App1}
\end{figure*}
Consider figure~\ref{fig:App1}. It is convenient to introduce the notation $\bracketed{\Delta x(j)}_1$ for 
\begin{equation}
\bracketed{\Delta x(j)}_1\equiv{1\over z} \sum_{<kj>} p_B(k) p_{acc}(k \to j)\Delta {\bf x}_{kj} \;,
\end{equation}
where $<kj>$ denotes that $k$ is a nearest neighbour of $j$. Then the contribution to the correlation function $\bracketed{\Delta {\bf x}_1\cdot\Delta {\bf x}_2}$ for site $1$ f is the weighted average over all neighbours of site $1$ of the product $\Delta {\bf x}_{1j}\cdot \bracketed{\Delta x(j)}_1$ (see Fig~\ref{fig:App1}(B)). 
Figure~\ref{fig:App1}(C) and (D) show examples of the subsequent steps. 

To obtain the contribution to the correlation function due to site $i$, we compute:
\begin{equation}
\bracketed{\Delta {\bf x}_1\cdot\Delta {\bf x}_2}=\sum_{i=1}^M{1\over z} \sum_{<ji>}p_{acc}(j\to i) \Delta {\bf x}_{ji}\cdot  \bracketed{\Delta {\bf x}(j)}_1
\end{equation}

Continuing the recursive procedure, we define $\bracketed{\Delta {\bf x}(j)}_2$ as the average of the one-step displacement vector $\bracketed{\Delta {\bf x}(k)}_1$ of all particles that can reach site $j$ in two steps from a site $k$. The contribution to the correlation function is then 
\begin{equation}
{1\over z} \sum_{j n.n i}p_{acc}(j\to i)\Delta {\bf x}_{ji}\cdot \bracketed{\Delta {\bf x}(j)}_2
\end{equation}
To obtain the desired correlation function by multiplying the propagated vectors with ${1\over z} p_{acc}(n-1 \to n)\Delta {\bf x}_{n-1,n}$, for all $z$ neighbours of site $n$.
In general, then, the expression for $\bracketed{\Delta {\bf x}_1\cdot\Delta {\bf x}_n}$ is
\begin{eqnarray}
\bracketed{\Delta {\bf x}_1\cdot\Delta {\bf x}_n} & = & \sum_{i=1}^M {1\over z} \sum_{<ji>}p_{acc}(j\to i) \\
& \times & \sum_k {\rm Prob}(k\to j;n-2) \Delta {\bf x}_{ji}\cdot \bracketed{\Delta x(k)}_1 \nonumber 
\end{eqnarray}
where ${\rm Prob}(k\to j;n-2) $ denotes the probability that a particle diffuses in $n-2$ steps from site $k$ to site $j$.
\subsection{NMR signal}\label{app:NMR}
In this case, we wish to compute
\begin{equation}  
G(t)=\langle {\rm e}^{{\rm i}\int_0^t \nu(t').dt'}\rangle
\end{equation} 
\newline
Again, time is discretised, i.e. $t=n\Delta t$. Then
\begin{equation}  
G(n\Delta t)=\langle {\rm e}^{{\rm i} \Delta t\sum_{n=1}^N \nu(n)}\rangle
\end{equation}
As before, we can compute this function recursively. We start by defining $g_i(\Delta t)$:
\begin{equation}
g_i(\Delta t) = p_B(i) e^{{\rm i} \Delta t \nu_i}\;.
\end{equation}
where $p_B(i)$ is the probability of finding a particle in site~$i$. Then
\begin{equation}
G(\Delta t) = \sum_{i=1}^M g_i(\Delta t)
\end{equation}
where $M$ is the number of lattice sites. To compute $G(2\Delta t)$, we use a recursive expression. We define $g_i(2\Delta t)$ as
\begin{equation}
g_i(2\Delta t)= e^{{\rm i} \Delta t \nu_i} \sum_{<ji>} p_{acc}(j\to i) g_j(\Delta t) 
\end{equation}
and
\begin{equation}
G(2\Delta t)=\sum_{i=1}^M g_i(2\Delta t) \;.
\end{equation}
In general
\begin{equation}\label{eq:NMRMP}
g_i(n\Delta t)=e^{{\rm i} \Delta t \nu_i} \sum_{<ji>} p_{acc}(j\to i) g_j((n-1)\Delta t) 
\end{equation}
and
\begin{equation}
G(n\Delta t)=\sum_{i=1}^M g_n(i) \;.
\end{equation}

The advantage of the Moment propagation method is that the computation effort scales as the number of lattice sites $\times$ the number of timesteps $N$, rather than as $z^N$.


\begin{thebibliography}{37}
\expandafter\ifx\csname natexlab\endcsname\relax\def\natexlab#1{#1}\fi
\expandafter\ifx\csname bibnamefont\endcsname\relax
  \def\bibnamefont#1{#1}\fi
\expandafter\ifx\csname bibfnamefont\endcsname\relax
  \def\bibfnamefont#1{#1}\fi
\expandafter\ifx\csname citenamefont\endcsname\relax
  \def\citenamefont#1{#1}\fi
\expandafter\ifx\csname url\endcsname\relax
  \def\url#1{\texttt{#1}}\fi
\expandafter\ifx\csname urlprefix\endcsname\relax\def\urlprefix{URL }\fi
\providecommand{\bibinfo}[2]{#2}
\providecommand{\eprint}[2][]{\url{#2}}

\bibitem[{\citenamefont{Zhao}(2006)}]{Zhao06}
\bibinfo{author}{\bibfnamefont{X.~S.} \bibnamefont{Zhao}}, \bibinfo{journal}{J.
  Mater. Chem.} \textbf{\bibinfo{volume}{16}}, \bibinfo{pages}{623}
  (\bibinfo{year}{2006}).

\bibitem[{\citenamefont{Xing et~al.}(2014)\citenamefont{Xing, Luo, Kleinhammes,
  and Wu}}]{Xing14}
\bibinfo{author}{\bibfnamefont{Y.-Z.} \bibnamefont{Xing}},
  \bibinfo{author}{\bibfnamefont{Z.-X.} \bibnamefont{Luo}},
  \bibinfo{author}{\bibfnamefont{A.}~\bibnamefont{Kleinhammes}},
  \bibnamefont{and} \bibinfo{author}{\bibfnamefont{Y.}~\bibnamefont{Wu}},
  \bibinfo{journal}{Carbon} \textbf{\bibinfo{volume}{77}},
  \bibinfo{pages}{1132} (\bibinfo{year}{2014}).

\bibitem[{\citenamefont{Wang et~al.}(2011)\citenamefont{Wang, K{\"o}ster,
  Trease, S{\'e}galini, Taberna, Simon, Gogotsi, and Grey}}]{Wang11b}
\bibinfo{author}{\bibfnamefont{H.}~\bibnamefont{Wang}},
  \bibinfo{author}{\bibfnamefont{T.~K.~J.} \bibnamefont{K{\"o}ster}},
  \bibinfo{author}{\bibfnamefont{N.~M.} \bibnamefont{Trease}},
  \bibinfo{author}{\bibfnamefont{J.}~\bibnamefont{S{\'e}galini}},
  \bibinfo{author}{\bibfnamefont{P.-L.} \bibnamefont{Taberna}},
  \bibinfo{author}{\bibfnamefont{P.}~\bibnamefont{Simon}},
  \bibinfo{author}{\bibfnamefont{Y.}~\bibnamefont{Gogotsi}}, \bibnamefont{and}
  \bibinfo{author}{\bibfnamefont{C.~P.} \bibnamefont{Grey}},
  \bibinfo{journal}{J. Am. Chem. Soc.} \textbf{\bibinfo{volume}{133}},
  \bibinfo{pages}{19270} (\bibinfo{year}{2011}).

\bibitem[{\citenamefont{Forse et~al.}(2013)\citenamefont{Forse, Griffin, Wang,
  Trease, Presser, Gogotsi, Simon, and Grey}}]{Forse13}
\bibinfo{author}{\bibfnamefont{A.~C.} \bibnamefont{Forse}},
  \bibinfo{author}{\bibfnamefont{J.~M.} \bibnamefont{Griffin}},
  \bibinfo{author}{\bibfnamefont{H.}~\bibnamefont{Wang}},
  \bibinfo{author}{\bibfnamefont{N.~M.} \bibnamefont{Trease}},
  \bibinfo{author}{\bibfnamefont{V.}~\bibnamefont{Presser}},
  \bibinfo{author}{\bibfnamefont{Y.}~\bibnamefont{Gogotsi}},
  \bibinfo{author}{\bibfnamefont{P.}~\bibnamefont{Simon}}, \bibnamefont{and}
  \bibinfo{author}{\bibfnamefont{C.~P.} \bibnamefont{Grey}},
  \bibinfo{journal}{Phys. Chem. Chem. Phys.} \textbf{\bibinfo{volume}{15}},
  \bibinfo{pages}{7722} (\bibinfo{year}{2013}).

\bibitem[{\citenamefont{Wang et~al.}(2013)\citenamefont{Wang, Forse, Griffin,
  Trease, Trognko, Taberna, Simon, and Grey}}]{Wang13}
\bibinfo{author}{\bibfnamefont{H.}~\bibnamefont{Wang}},
  \bibinfo{author}{\bibfnamefont{A.~C.} \bibnamefont{Forse}},
  \bibinfo{author}{\bibfnamefont{J.~M.} \bibnamefont{Griffin}},
  \bibinfo{author}{\bibfnamefont{N.~M.} \bibnamefont{Trease}},
  \bibinfo{author}{\bibfnamefont{L.}~\bibnamefont{Trognko}},
  \bibinfo{author}{\bibfnamefont{P.-L.} \bibnamefont{Taberna}},
  \bibinfo{author}{\bibfnamefont{P.}~\bibnamefont{Simon}}, \bibnamefont{and}
  \bibinfo{author}{\bibfnamefont{C.~P.} \bibnamefont{Grey}},
  \bibinfo{journal}{J. Am. Chem. Soc.} \textbf{\bibinfo{volume}{135}},
  \bibinfo{pages}{18968} (\bibinfo{year}{2013}).

\bibitem[{\citenamefont{Lazzeretti}(2000)}]{Lazzeretti00}
\bibinfo{author}{\bibfnamefont{P.}~\bibnamefont{Lazzeretti}},
  \bibinfo{journal}{Prog. Nucl. Mag. Res. Sp.} \textbf{\bibinfo{volume}{36}},
  \bibinfo{pages}{1 } (\bibinfo{year}{2000}). 

\bibitem[{\citenamefont{Forse et~al.}(2014)\citenamefont{Forse, Griffin,
  Presser, Gogotsi, and Grey}}]{Forse14}
\bibinfo{author}{\bibfnamefont{A.~C.} \bibnamefont{Forse}},
  \bibinfo{author}{\bibfnamefont{J.~M.} \bibnamefont{Griffin}},
  \bibinfo{author}{\bibfnamefont{V.}~\bibnamefont{Presser}},
  \bibinfo{author}{\bibfnamefont{Y.}~\bibnamefont{Gogotsi}}, \bibnamefont{and}
  \bibinfo{author}{\bibfnamefont{C.~P.} \bibnamefont{Grey}},
  \bibinfo{journal}{J. Phys. Chem. C} \textbf{\bibinfo{volume}{118}},
  \bibinfo{pages}{7508} (\bibinfo{year}{2014}).

\bibitem[{\citenamefont{Borchardt et~al.}(2013)\citenamefont{Borchardt,
  Oschatz, Paasch, Kaskel, and Brunner}}]{Borchardt13}
\bibinfo{author}{\bibfnamefont{L.}~\bibnamefont{Borchardt}},
  \bibinfo{author}{\bibfnamefont{M.}~\bibnamefont{Oschatz}},
  \bibinfo{author}{\bibfnamefont{S.}~\bibnamefont{Paasch}},
  \bibinfo{author}{\bibfnamefont{S.}~\bibnamefont{Kaskel}}, \bibnamefont{and}
  \bibinfo{author}{\bibfnamefont{E.}~\bibnamefont{Brunner}},
  \bibinfo{journal}{Phys. Chem. Chem. Phys.} \textbf{\bibinfo{volume}{15}},
  \bibinfo{pages}{15177} (\bibinfo{year}{2013}).

\bibitem[{\citenamefont{Anderson et~al.}(2010)\citenamefont{Anderson,
  McNicholas, Kleinhammes, Wang, Liu, and Wu}}]{Anderson10}
\bibinfo{author}{\bibfnamefont{R.~J.} \bibnamefont{Anderson}},
  \bibinfo{author}{\bibfnamefont{T.~P.} \bibnamefont{McNicholas}},
  \bibinfo{author}{\bibfnamefont{A.}~\bibnamefont{Kleinhammes}},
  \bibinfo{author}{\bibfnamefont{A.}~\bibnamefont{Wang}},
  \bibinfo{author}{\bibfnamefont{J.}~\bibnamefont{Liu}}, \bibnamefont{and}
  \bibinfo{author}{\bibfnamefont{Y.}~\bibnamefont{Wu}}, \bibinfo{journal}{J.
  Am. Chem. Soc.} \textbf{\bibinfo{volume}{132}}, \bibinfo{pages}{8618}
  (\bibinfo{year}{2010}).

\bibitem[{\citenamefont{Haddon}(1995)}]{Haddon95}
\bibinfo{author}{\bibfnamefont{R.~C.} \bibnamefont{Haddon}},
  \bibinfo{journal}{Nature} \textbf{\bibinfo{volume}{378}},
  \bibinfo{pages}{249} (\bibinfo{year}{1995}).

\bibitem[{\citenamefont{Heine et~al.}(2005)\citenamefont{Heine, Corminboeuf,
  and Seifert}}]{Heine05}
\bibinfo{author}{\bibfnamefont{T.}~\bibnamefont{Heine}},
  \bibinfo{author}{\bibfnamefont{C.}~\bibnamefont{Corminboeuf}},
  \bibnamefont{and} \bibinfo{author}{\bibfnamefont{G.}~\bibnamefont{Seifert}},
  \bibinfo{journal}{Chem. Rev.} \textbf{\bibinfo{volume}{105}},
  \bibinfo{pages}{3889} (\bibinfo{year}{2005}).

\bibitem[{\citenamefont{Facelli}(2006)}]{Facelli06}
\bibinfo{author}{\bibfnamefont{J.}~\bibnamefont{Facelli}},
  \bibinfo{journal}{Magn. Reson. Chem.} \textbf{\bibinfo{volume}{44}},
  \bibinfo{pages}{401} (\bibinfo{year}{2006}).

\bibitem[{\citenamefont{Levesque et~al.}(2013)\citenamefont{Levesque, Duvail,
  Pagonabarraga, Frenkel, and Rotenberg}}]{Levesque13}
\bibinfo{author}{\bibfnamefont{M.}~\bibnamefont{Levesque}},
  \bibinfo{author}{\bibfnamefont{M.}~\bibnamefont{Duvail}},
  \bibinfo{author}{\bibfnamefont{I.}~\bibnamefont{Pagonabarraga}},
  \bibinfo{author}{\bibfnamefont{D.}~\bibnamefont{Frenkel}}, \bibnamefont{and}
  \bibinfo{author}{\bibfnamefont{B.}~\bibnamefont{Rotenberg}},
  \bibinfo{journal}{Phys. Rev. E} \textbf{\bibinfo{volume}{88}},
  \bibinfo{pages}{013308} (\bibinfo{year}{2013}).

\bibitem[{\citenamefont{Sen}(2003)}]{Sen03}
\bibinfo{author}{\bibfnamefont{P.~N.} \bibnamefont{Sen}}, \bibinfo{journal}{J.
  Chem. Phys.} \textbf{\bibinfo{volume}{119}}, \bibinfo{pages}{9871}
  (\bibinfo{year}{2003}).

\bibitem[{\citenamefont{Sen}(2004)}]{Sen04}
\bibinfo{author}{\bibfnamefont{P.~N.} \bibnamefont{Sen}},
  \bibinfo{journal}{Concept Magnetic Res. Part A}
  \textbf{\bibinfo{volume}{23A}}, \bibinfo{pages}{1} (\bibinfo{year}{2004}).

\bibitem[{\citenamefont{Dudko et~al.}(2005)\citenamefont{Dudko, Berezhkovskii,
  and Weiss}}]{Dudko05}
\bibinfo{author}{\bibfnamefont{O.~K.} \bibnamefont{Dudko}},
  \bibinfo{author}{\bibfnamefont{A.~M.} \bibnamefont{Berezhkovskii}},
  \bibnamefont{and} \bibinfo{author}{\bibfnamefont{G.~H.} \bibnamefont{Weiss}},
  \bibinfo{journal}{J. Phys. Chem. B} \textbf{\bibinfo{volume}{109}},
  \bibinfo{pages}{21296} (\bibinfo{year}{2005}).

\bibitem[{\citenamefont{Ravikovitch and Neimark}(2001)}]{Ravikovitch01}
\bibinfo{author}{\bibfnamefont{P.~I.} \bibnamefont{Ravikovitch}}
  \bibnamefont{and} \bibinfo{author}{\bibfnamefont{A.~V.}
  \bibnamefont{Neimark}}, \bibinfo{journal}{Colloids and Surfaces A:
  Physicochemical and Engineering Aspects} \textbf{\bibinfo{volume}{187--188}},
  \bibinfo{pages}{11} (\bibinfo{year}{2001}).

\bibitem[{\citenamefont{Dash et~al.}(2006)\citenamefont{Dash, Chmiola, Yushin,
  Gogotsi, Laudisio, Singer, Fischer, and Kucheyev}}]{Dash06}
\bibinfo{author}{\bibfnamefont{R.}~\bibnamefont{Dash}},
  \bibinfo{author}{\bibfnamefont{J.}~\bibnamefont{Chmiola}},
  \bibinfo{author}{\bibfnamefont{G.}~\bibnamefont{Yushin}},
  \bibinfo{author}{\bibfnamefont{Y.}~\bibnamefont{Gogotsi}},
  \bibinfo{author}{\bibfnamefont{G.}~\bibnamefont{Laudisio}},
  \bibinfo{author}{\bibfnamefont{J.}~\bibnamefont{Singer}},
  \bibinfo{author}{\bibfnamefont{J.}~\bibnamefont{Fischer}}, \bibnamefont{and}
  \bibinfo{author}{\bibfnamefont{S.}~\bibnamefont{Kucheyev}},
  \bibinfo{journal}{Carbon} \textbf{\bibinfo{volume}{44}},
  \bibinfo{pages}{2489} (\bibinfo{year}{2006}).

\bibitem[{\citenamefont{Feng et~al.}(2010{\natexlab{a}})\citenamefont{Feng,
  Qiao, Huang, Sumpter, and Meunier}}]{Feng10a}
\bibinfo{author}{\bibfnamefont{G.}~\bibnamefont{Feng}},
  \bibinfo{author}{\bibfnamefont{R.}~\bibnamefont{Qiao}},
  \bibinfo{author}{\bibfnamefont{J.}~\bibnamefont{Huang}},
  \bibinfo{author}{\bibfnamefont{B.~G.} \bibnamefont{Sumpter}},
  \bibnamefont{and} \bibinfo{author}{\bibfnamefont{V.}~\bibnamefont{Meunier}},
  \bibinfo{journal}{J. Phys. Chem. C} \textbf{\bibinfo{volume}{114}},
  \bibinfo{pages}{18012} (\bibinfo{year}{2010}{\natexlab{a}}).

\bibitem[{\citenamefont{Feng et~al.}(2010{\natexlab{b}})\citenamefont{Feng,
  Qiao, Huang, Sumpter, and Meunier}}]{Feng10c}
\bibinfo{author}{\bibfnamefont{G.}~\bibnamefont{Feng}},
  \bibinfo{author}{\bibfnamefont{R.}~\bibnamefont{Qiao}},
  \bibinfo{author}{\bibfnamefont{J.}~\bibnamefont{Huang}},
  \bibinfo{author}{\bibfnamefont{B.~G.} \bibnamefont{Sumpter}},
  \bibnamefont{and} \bibinfo{author}{\bibfnamefont{V.}~\bibnamefont{Meunier}},
  \bibinfo{journal}{ACS Nano} \textbf{\bibinfo{volume}{4}},
  \bibinfo{pages}{2382} (\bibinfo{year}{2010}{\natexlab{b}}).

\bibitem[{\citenamefont{Xing et~al.}(2013)\citenamefont{Xing, Vatamanu,
  Borodin, and Bedrov}}]{Xing13}
\bibinfo{author}{\bibfnamefont{L.}~\bibnamefont{Xing}},
  \bibinfo{author}{\bibfnamefont{J.}~\bibnamefont{Vatamanu}},
  \bibinfo{author}{\bibfnamefont{O.}~\bibnamefont{Borodin}}, \bibnamefont{and}
  \bibinfo{author}{\bibfnamefont{D.}~\bibnamefont{Bedrov}},
  \bibinfo{journal}{J. Phys. Chem. Lett.} \textbf{\bibinfo{volume}{4}},
  \bibinfo{pages}{132} (\bibinfo{year}{2013}).

\bibitem[{\citenamefont{Jiang et~al.}(2012)\citenamefont{Jiang, Jin, Henderson,
  and Wu}}]{Jiang12b}
\bibinfo{author}{\bibfnamefont{D.-e.} \bibnamefont{Jiang}},
  \bibinfo{author}{\bibfnamefont{Z.}~\bibnamefont{Jin}},
  \bibinfo{author}{\bibfnamefont{D.}~\bibnamefont{Henderson}},
  \bibnamefont{and} \bibinfo{author}{\bibfnamefont{J.}~\bibnamefont{Wu}},
  \bibinfo{journal}{J. Phys. Chem. Lett.} \textbf{\bibinfo{volume}{3}},
  \bibinfo{pages}{1727} (\bibinfo{year}{2012}).

\bibitem[{\citenamefont{Frenkel}(1987)}]{Frenkel87}
\bibinfo{author}{\bibfnamefont{D.}~\bibnamefont{Frenkel}},
  \bibinfo{journal}{Phys. Lett. A} \textbf{\bibinfo{volume}{121}},
  \bibinfo{pages}{385} (\bibinfo{year}{1987}).

\bibitem[{\citenamefont{Rotenberg et~al.}(2008)\citenamefont{Rotenberg,
  Pagonabarraga, and Frenkel}}]{Rotenberg08}
\bibinfo{author}{\bibfnamefont{B.}~\bibnamefont{Rotenberg}},
  \bibinfo{author}{\bibfnamefont{I.}~\bibnamefont{Pagonabarraga}},
  \bibnamefont{and} \bibinfo{author}{\bibfnamefont{D.}~\bibnamefont{Frenkel}},
  \bibinfo{journal}{Europhys. Lett.} \textbf{\bibinfo{volume}{83}},
  \bibinfo{pages}{34004} (\bibinfo{year}{2008}).

\bibitem[{\citenamefont{Levitt}(2008)}]{Levitt_book}
\bibinfo{author}{\bibfnamefont{M.~H.} \bibnamefont{Levitt}},
  \emph{\bibinfo{title}{Spin dynamics, Basis of nuclear magnetic resonance}}
  (\bibinfo{publisher}{John Wiley and Sons, Ltd}, \bibinfo{year}{2008}).

\bibitem[{\citenamefont{Cavanagh et~al.}(1996)\citenamefont{Cavanagh,
  Fairbrother, Palmer~III, and Skelton}}]{Cavanagh_book}
\bibinfo{author}{\bibfnamefont{J.}~\bibnamefont{Cavanagh}},
  \bibinfo{author}{\bibfnamefont{W.~J.} \bibnamefont{Fairbrother}},
  \bibinfo{author}{\bibfnamefont{A.~G.} \bibnamefont{Palmer~III}},
  \bibnamefont{and} \bibinfo{author}{\bibfnamefont{N.~J.}
  \bibnamefont{Skelton}}, \emph{\bibinfo{title}{Protein NMR spectroscopy,
  Principles and practice}} (\bibinfo{publisher}{Academic Press, Inc.},
  \bibinfo{year}{1996}).

\bibitem[{\citenamefont{Merlet et~al.}(2013)\citenamefont{Merlet, Salanne,
  Rotenberg, and Madden}}]{Merlet13b}
\bibinfo{author}{\bibfnamefont{C.}~\bibnamefont{Merlet}},
  \bibinfo{author}{\bibfnamefont{M.}~\bibnamefont{Salanne}},
  \bibinfo{author}{\bibfnamefont{B.}~\bibnamefont{Rotenberg}},
  \bibnamefont{and} \bibinfo{author}{\bibfnamefont{P.~A.}
  \bibnamefont{Madden}}, \bibinfo{journal}{Electrochimica Acta}
  \textbf{\bibinfo{volume}{101}}, \bibinfo{pages}{262} (\bibinfo{year}{2013}).

\bibitem[{\citenamefont{Juselius and Sundholm}(1999)}]{Juselius99}
\bibinfo{author}{\bibfnamefont{J.}~\bibnamefont{Juselius}} \bibnamefont{and}
  \bibinfo{author}{\bibfnamefont{D.}~\bibnamefont{Sundholm}},
  \bibinfo{journal}{Phys. Chem. Chem. Phys.} \textbf{\bibinfo{volume}{1}},
  \bibinfo{pages}{3429} (\bibinfo{year}{1999}).

\bibitem[{\citenamefont{Merlet}(2013)}]{These_Merlet}
\bibinfo{author}{\bibfnamefont{C.}~\bibnamefont{Merlet}}, Ph.D. thesis,
  \bibinfo{school}{Universit{\'e} Pierre et Marie Curie}
  (\bibinfo{year}{2013}).

\bibitem[{\citenamefont{Forse et~al.}()\citenamefont{Forse, Griffin, Merlet, Bayley
  Wang, Simon, and Grey}}]{Forse14b}
\bibinfo{author}{\bibfnamefont{A.~C.} \bibnamefont{Forse}},
  \bibinfo{author}{\bibfnamefont{J.~M.} \bibnamefont{Griffin}},
  \bibinfo{author}{\bibfnamefont{C.}~\bibnamefont{Merlet}},
  \bibinfo{author}{\bibfnamefont{P.~M.}~\bibnamefont{Bayley}},
  \bibinfo{author}{\bibfnamefont{H.}~\bibnamefont{Wang}},
  \bibinfo{author}{\bibfnamefont{P.}~\bibnamefont{Simon}}, \bibnamefont{and}
  \bibinfo{author}{\bibfnamefont{C.~P.} \bibnamefont{Grey}},
  \bibinfo{journal}{in preparation}.

\end{thebibliography}

\end{document}